\documentclass[runningheads]{llncs}
\usepackage[T1]{fontenc}
\usepackage{graphicx}
\usepackage{booktabs}
\usepackage{caption}
\usepackage{adjustbox}

\usepackage{booktabs}
\usepackage{caption}
\usepackage{adjustbox}
\usepackage[most]{tcolorbox}
\usepackage{listings}
\usepackage{xcolor}
\lstdefinelanguage{gherkin}{
  morekeywords={Scenario, Given, When, Then, And, But},
  sensitive=true,
}

\usepackage{listings}
\usepackage{xcolor}

\lstdefinelanguage{gherkin}{
  morekeywords={Scenario, Given, When, Then, And, But},
  sensitive=true
}

\lstdefinestyle{gherkin-style}{
  language=gherkin,
  basicstyle=\ttfamily\small,
  keywordstyle=\color{blue}\bfseries,   
  frame=single,
  breaklines=true,
  backgroundcolor=\color{gray!5},
  columns=fullflexible,
  keepspaces=true,
  captionpos=b,
  moredelim=[is][\color{gray}\itshape]{[}{]},   
  literate=
    {Vehicle.Cabin.ChildPresenceDetection.IsChildDetected}{{\color{red}Vehicle.Cabin.ChildPresenceDetection.IsChildDetected}}1
    {Vehicle.Cabin.Infotainment.HVAC.AutoOverrideActive}{{\color{red}Vehicle.Cabin.Infotainment.HVAC.AutoOverrideActive}}1
}

\begin{document}
\title{GenAI for Automotive Software Development: From Requirements to Wheels}
%
%
\author{Nenad Petrovic\inst{1}\orcidID{0000-0003-2264-7369} \and
 Fengjunjie Pan\inst{1}\orcidID{0009-0005-8303-1156} \and Vahid Zolfaghari\inst{1}\orcidID{0009-0004-0039-6014} \and Krzysztof Lebioda\inst{1}\orcidID{0000-0002-7905-8103} \and Andre Schamschurko\inst{1}\orcidID{0009-0000-7030-0955} \and Alois Knoll\inst{1}\orcidID{0000-0003-4840-076X}}
\authorrunning{N. Petrovic et al.}
\institute{Technical University of Munich, Germany
\email{nenad.petrovic@tum.de,f.pan@tum.de,v.zolfaghari@tum.de, krzysztof.lebioda@tum.de, andre.schamschurko@tum.de, k@tum.de}\\
}

%
\maketitle              
\begin{abstract}
This paper introduces a GenAI-empowered approach to automated development of automotive software, with emphasis on autonomous and Advanced Driver Assistance Systems (ADAS) capabilities. The process starts with requirements as input, while the main generated outputs are test scenario code for simulation environment, together with implementation of desired ADAS capabilities targeting hardware platform of the vehicle connected to testbench. Moreover, we introduce additional steps for requirements consistency checking leveraging Model-Driven Engineering (MDE). In the proposed workflow, Large Language Models (LLMs) are used for model-based summarization of requirements (Ecore metamodel, XMI model instance and OCL constraint creation), test scenario generation, simulation code (Python) and target platform code generation (C++). Additionally, Retrieval Augmented Generation (RAG) is adopted to enhance test scenario generation from autonomous driving regulations-related documents. Our approach aims shorter compliance and re-engineering cycles, as well as reduced development and testing time when it comes to ADAS-related capabilities.

\end{abstract}
\section{Introduction}
Automotive industry is characterized by strict design, development, testing, and manufacturing processes that must comply with a variety of regulations and standards. These constraints often result in lengthy timelines from research and development to full-scale production. Innovation in this field is further challenged by time-intensive, costly procedures that depend heavily on specialized domain expertise and manual effort. Considering the increasing complexity of Software Defined Vehicle (SDV) products incorporating autonomous and assisted driving capabilities over years \cite{McKinsey2020}, the development of even standard-sized vehicles involves managing hundreds of thousands of requirements, which introduces additional gap between development and production. 

Artificial Intelligence (AI), especially the novel GenAI-based solutions exhibit strong potential when it comes to bridging this gap and automotive software development process automation. Current research indicates that the adoption of GenAI in automotive software development primarily focuses on areas such as requirements management, compliance, test scenario generation, and code generation. Nevertheless, implementing GenAI in sensitive industries like automotive brings significant challenges \cite{staron2025} \cite{phatale2024}. A key concern is that GenAI models are prone to hallucinations—producing plausible but inaccurate or fabricated information. This inherent uncertainty makes the direct use of unverified AI outputs impractical, so additional validation steps through formally grounded methods such as Model-Driven Engineering (MDE) are identified as a possible solution to tackle them. On the other side, there is a growing need for smaller, locally deployable AI models tailored to specific, narrowly defined tasks, due to potential constrains of exposing automotive software assets (requirements and code).

In this paper, we consider the adoption of state-of-art GenAI models and techniques - particularly Large Language Models (LLM) and Retrieval Augmented Generation (RAG), considering the complete software development workflow - from requirements and compliance documents to tests and target platform code generation. Additionally, we leverage MDE and hallucination tackling techniques to in order to increase trustworthiness and reproducibility of the approach.

\section{Methodology}
\label{method}
In this section, we present a workflow for a GenAI-powered approach to automotive software development from end to end, building upon our previous works \cite{petrovic2024synergy}. The proposed workflow is depicted in Fig. 1, where we indicate step automation by AI, along with auxiliary techniques (such as MDE). The workflow begins with the processing of input documents—such as customer-specific requirements or regulatory standards—using Retrieval-Augmented Generation (RAG). This step facilitates efficient extraction of relevant information, which is then used to build datasets for code and test generation. In this context, regulatory documents like UN152 \cite{unece2020reg152} often serve as the foundation for generating test scenarios via RAG. Since visual elements (e.g., diagrams, graphs) in automotive documentation can convey crucial information, Vision-Language Models (VLMs) may also be used to extract such content \cite{petrovic2025multimodal}. The next key step involves creating a formal representation, which serves as an intermediary between requirements extraction and code generation. Large Language Models (LLMs) are employed to summarize and structure the extracted requirements into a formal template or metamodel, enabling early design-time checks such as compliance verification. The metamodel itself can be manually crafted or constructed automatically. Once the requirements are checked for completeness, correctness, and compliance, code generation can proceed using LLMs. Based on extracted test scenario, the proposed vehicle configuration is first assessed in simulation environment (Python code in CARLA). Afterwards, code for target platform (vehicle mounted on the testbench and connected to simulation environment) is generated, together with complementary simulation code. While it is still recommended to include human reviewer in a loop, we try to reduce the need for manual intervention through the use of multi-agent LLM systems for tackling the possible hallucinations \cite{schamschurko2025recsip}.

\begin{figure}
    \centering
    \includegraphics[width=1\linewidth]{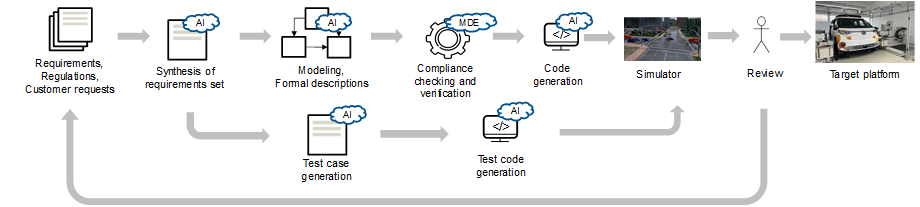}
    \caption{GenAI-empowered automotive software development workflow.}
    \label{fig:genaiwf}
\end{figure}

\section{Implementation}
In this section, we provide insights into implementation of distinct workflow steps. Their integration was done relying on n8n workflow automation tool \cite{n8n2025}. 
\label{implementation}

\subsection{Model Checker} 
\begin{figure}
    \centering
    \includegraphics[width=1\linewidth]{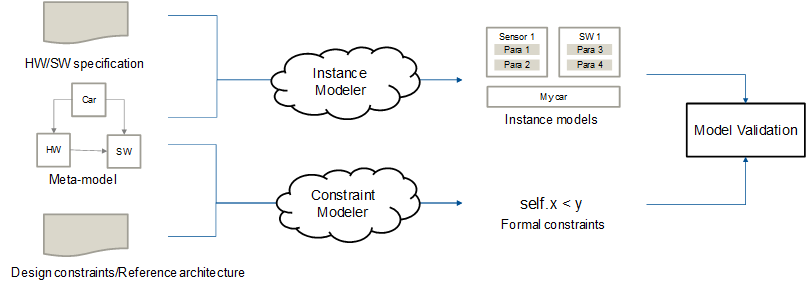}
    \caption{Model instance and OCL rule generation for consistency checking.}
    \label{fig:modelai}
\end{figure}

Model Checker component consists of two LLM agents: 1) model instance generation 2) constraint generation. The aim of this solution is to enable the adoption of model-driven approach for consistency checking, where set of Object Constraint Language Rule (OCL) is checked if satisfied within given XMI model instance., as depicted in Fig. \ref{fig:modelai}.
For the first task, we take HW/SW specifications and system's meta-model as input and create a XMI instance model representing the target system. We propose Llama 3.1-70B based solution that achieves a comparable semantic score to GPT-4o. Instead of direct XMI generation, our method makes use of simpler conceptual model notation as intermediate step in order to improve performance \cite{pan2025llmenabled}. 
On the other side, for OCL rule generation, we adopt custom, fine-tuned model based on custom-tailored Llama3-8B \cite{pan2024ocl} in synergy with RAG \cite{li2025optimizing}. The input of OCL generation is meta-model and design constraint, e.g., from the reference architecture. Solution is locally deployable.

While metamodel itself capturing the main aspects of the automotive system is usually output of manual design efforts, we also introduce LLM-driven approach aiming to perform this step automatically as well, incorporating implicit architecture formalization based on requirements. For this purpose we adopt locally deployable deepseek-ai/deepseek-llm-7b-chat \cite{deepseek2024llm7bchat}. Simpler PlantUML notation for metamodel representation is used, as well as iterative approach where smaller subset of requirements is provided as input in single step. Additionally, human reviewer has insight into visual representation of the metamodel and can provide feedback/corrections as input. Once human reviewer is satisfied, model-to-model transformation will be executed in order to construct Ecore metamodel from PlantUML, as shown in Fig. \ref{fig:metamodelai}. 
\begin{figure}
    \centering
    \includegraphics[width=1\linewidth]{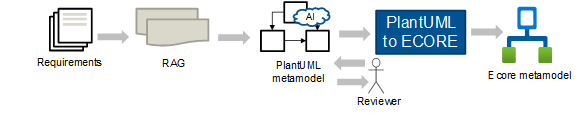}
    \caption{LLM-based iterative metamodeling.}
    \label{fig:metamodelai}
\end{figure}
\subsection{Regulation-Compliant Scenario Generation} 
\begin{figure}
    \centering
    \includegraphics[width=1\linewidth]{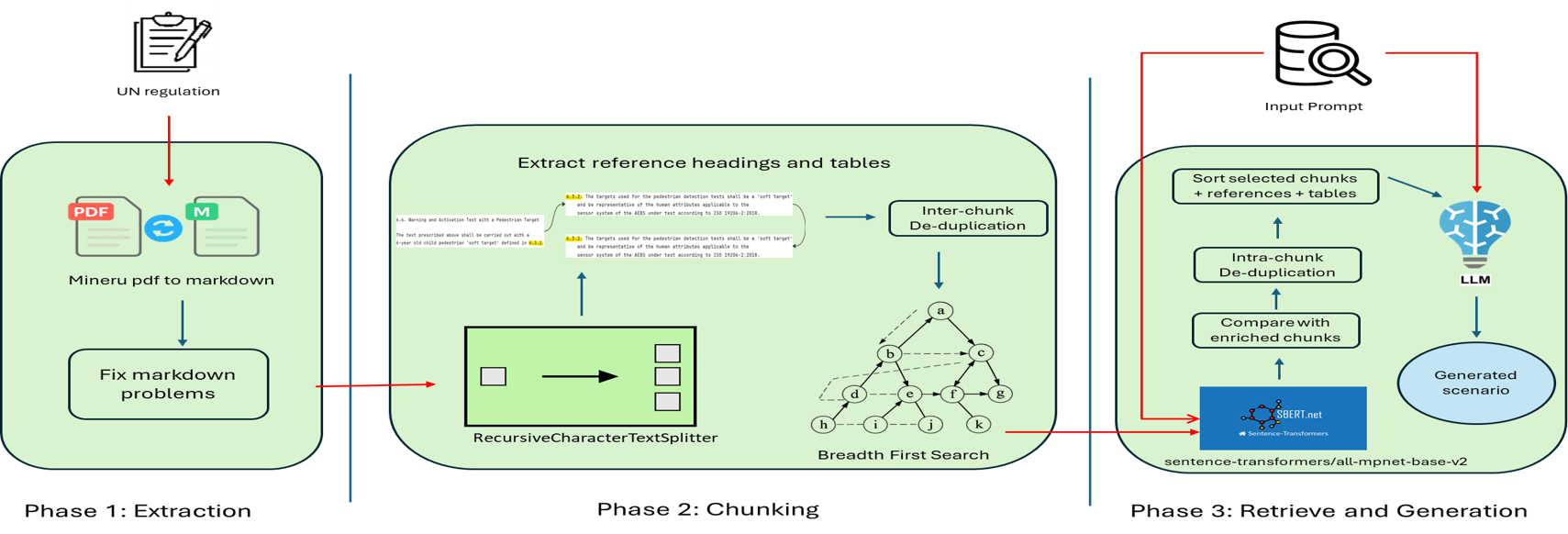}
    \caption{GenAI-empowered automotive software development workflow.}
    \label{fig:ragai}
\end{figure}

In automotive software development, especially for ADAS and Autonomous Driving Systems (ADS), producing precise, regulation-compliant test scenarios is essential \cite{zolfaghari2024rag}. Conventional manual methods for interpreting and verifying compliance with standards such as UN Regulation No. 152 are often inefficient, error-prone, and expensive. While Large Language Models (LLMs) can help automate this work, they frequently miss critical numerical details, fail to distinguish between different test conditions (like laden versus unladen vehicles), and struggle to process lengthy documents that exceed context window. Supplying the entire regulation to an LLM also consumes excessive tokens, driving up costs. The aim of this step is extracting test scenarios in textual format from standards like UN Regulation No. 152, focused on automated emergency braking systems (AEBS). Our RAG system uses a robust two-stage pipeline, as depicted in Fig. \ref{fig:ragai}. It applies a SmartChunking approach to preprocess PDF-based standards by mapping hierarchical paragraph structures, resolving nested references, and expanding chunks through graph-based traversal, ensuring token spending efficiency at the same time. Smart Retrieve and Rerank module then performs query-sensitive retrieval across these enriched chunks, leading to verifiable and more accurate results than conventional chunking or non-RAG setups. 

\subsection{Simulation Test Scenario Generation} 
\begin{figure}
    \centering
    \includegraphics[width=1\linewidth]{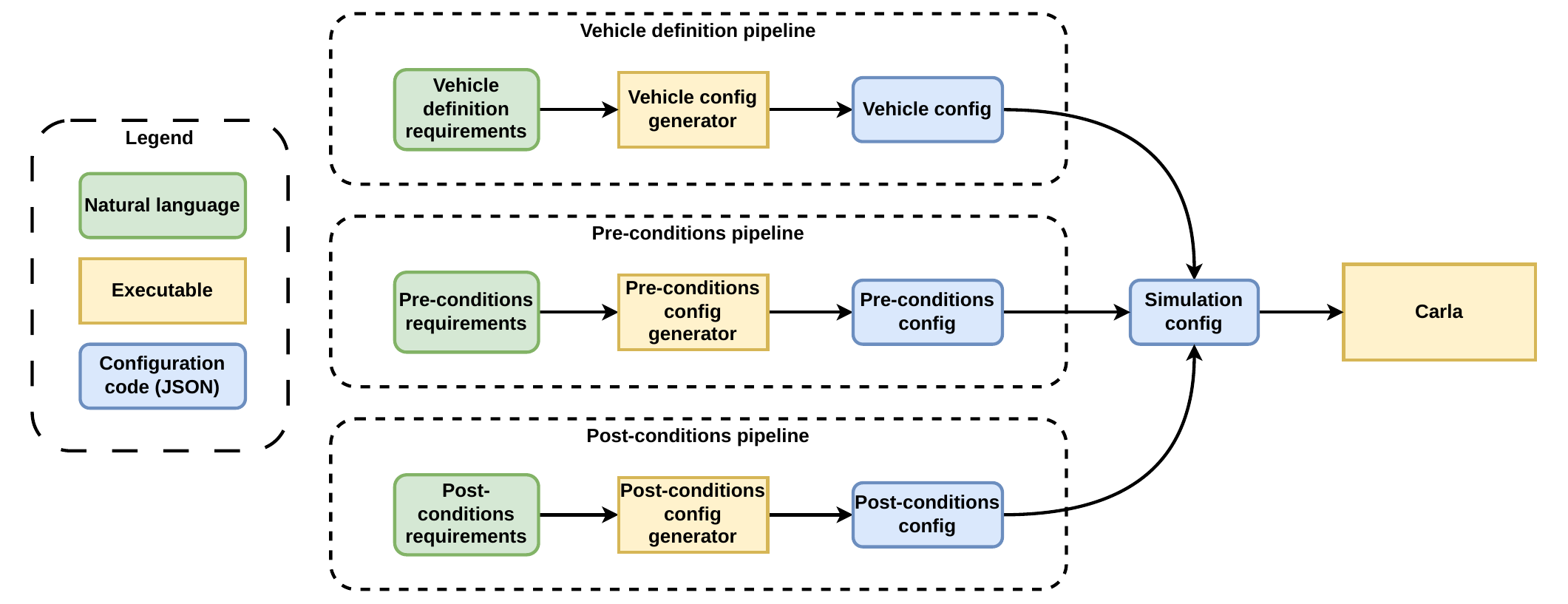}
    \caption{Simulation scenario generation.}
    \label{fig:carlai}
\end{figure}

Using regulation-compliant scenarios derived from UN Regulation No. 152 \cite{unece2020reg152} and similar sources, this component generates configuration code for a CARLA-based simulation environment.
Separate LLM-driven pipelines are employed to address the following requirement categories: 1) vehicle definition (including sensor specifications), 2) pre-conditions (such as scene setup, agent positioning, and weather settings), and 3) post-conditions (including telemetry data and expected simulation outcomes).
The approach uses GPT-4o and builds upon the methodology presented in \cite{lebioda2025requirements}.
The complete workflow is presented in Fig. \ref{fig:carlai}.

\subsection{Target Platform Code Generation} 

In the last step, C++ code for target testbench platform is also generated (relying on GPT-4o), while its role is complementary to Python simulation code executed in CARLA, as extension of our work from \cite{pan2025automating}. While sensing part is performed within simulation on CARLA server, control (steering, braking, acceleration) is executed on vehicle controller relying on comAPI (alternatively TC4D or CAN FD API). Events from simulation environment are sent from Python script which acts as ROS2 publisher, while C++ target platform code receives them as subscriber. Based on the received events, vehicle commands are executed. The factors taken into account for code generation are: 1) experiment model - contains the information about both the scenario and vehicle configuration 2) code templates 3) Vehicle Signal Specification (VSS) \cite{covesa_vss_2025} catalog - list of available vehicle signals. Before code generation, corresponding vehicle signals are mapped based on provided experiment description with respect to signal catalog, so the list of used signals is established. After that, control logic is generated, leveraging comAPI invocations based on VSS signals which are translated to CAN messages for zone ECUs thanks to gateway component.
\begin{figure}
    \centering
    \includegraphics[width=1\linewidth]{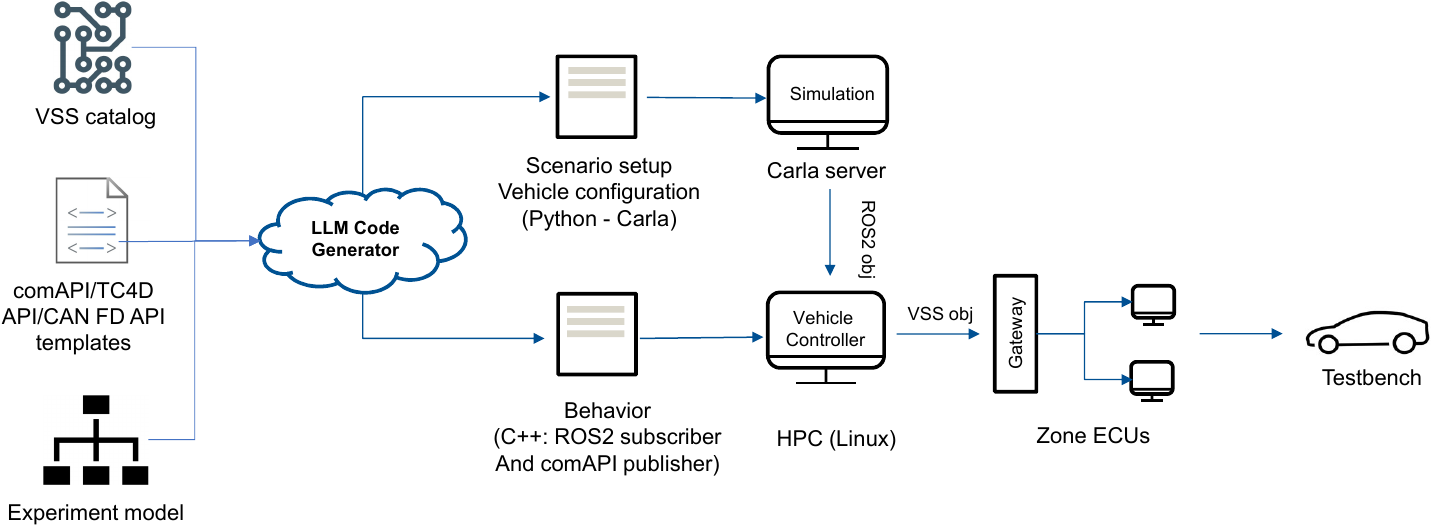}
    \caption{Target platform code generation.}
    \label{fig:metamodelai}
\end{figure}

\section{Conclusion}
Based on the initial results for automated emergency braking, the proposed approach leveraging GenAI for automotive engineering and re-engineering processes automation in order to tackle the cognitive load and increasing SDV systems complexity, exhibits strong potential for reducing the time needed for innovation and testing - from days and hours to order of magnitude of a minute.
\begin{credits}
\subsubsection{\ackname} This research was funded by the Federal Ministry of Research, Technology and Space of Germany as part of the CeCaS project, FKZ: 16ME0800K.
\end{credits}
%
%
%
\bibliographystyle{splncs04}
\bibliography{biblio}

\end{document}